\documentclass[useAMS,usenatbib,usegraphicx]{mn2e}

\title{Thermally Pulsing Asymptotic Giant Branch Star Models and
  Globular Cluster Planetary Nebulae I: The Model}
\author[J. F. Buell]{J. F. Buell$^{1}$\thanks{E-mail:
    buelljf@alfredstate.edu}\\ $^{1}$100 College Drive, SUNY College
  of Technology at Alfred, Alfred, NY 14843, USA\\ }
\begin{document}

\maketitle

\date{Accepted 2011 September 21. Received 2011 September 21; in
  original form 2011 August 8}

\pagerange{\pageref{firstpage}--\pageref{lastpage}} \pubyear{2011}

\label{firstpage}

\begin{abstract}
Thermally pulsing asymptotic giant branch models of globular cluster
stars are calculated using a synthetic model with the goal of
reproducing the chemical composition, core masses and other
observational parameters of the four known globular cluster planetary
nebulae as well as roughly matching the overall cluster
properties. The evolution of stars with an enhanced helium abundance
($Y$) and blue stragglers are modeled. New pre-thermally pulsing
asymptotic giant branch mass-losses for red giant branch and early
asymptotic giant branch stars are calculated from the Padova stellar
evolution models \citep{berta,bertb}. The new mass-losses are
calculated to get the relative differences in mass-losses due to
enhanced helium abundances.

The global properties of the globular cluster planetary nebula are
reproduced with these models. The metallicity, mass of the central
star, overall metallicities, helium abundance and the nebular mass are
matched to the observational values. Globular cluster planetary
nebulae JaFu 1 and JaFu 2 are reproduced {\it by assuming progenitor
  stars} with masses near the typical main sequence turn-offs of
globular clusters and with enhanced helium abundances very similar to
the enhancements inferred from fitting isochrones to globular cluster
colour-magnitude diagrams. The globular cluster PN GJJC-1 can be
roughly fit by a progenitor star with very extreme helium enhancement
($Y\approx0.40$) near the turn-off producing a central star with the
same mass as inferred by observations and a very low nebular mass. The
abundances and core mass of planetary nebula Ps 1 and its central star
(K648) are reproduced by a blue straggler model. However, it turned
out to be impossible to reproduce its nebular mass and it is concluded
some kind of binary scenario may be needed to explain K648.

\end{abstract}

\section{Introduction}
The globular cluster (GC) system of the Galaxy contains four known
planetary nebulae (PNe) all of which have unusual abundances. The
nebulae Ps 1 and its central star K648 in M15 have a very high C/O
ratio of nearly 10 \citep{al,bione,bitwo,how,rau}. The PNe JaFu 1 in
Pal 6 and JaFu 2 in NGC 6441 have high ratios of He/H \citep{jac}
(0.115 and 0.141 for JaFu 1 and JaFu 2, respectively). The fourth GC PN
GJJC-1 in M22 is one of three known PN with no hydrogen in the nebula
\citep{jac}. All four are very different from the typical disc
PN. They are even more difficult to explain since the typical age of a
globular cluster is $\sim12.5\,{\rm Gyr}$ leading to a turn-off mass
of $\sim0.85\,{\rm M}_{\sun}$. At this mass only minor changes from
the zero-age main sequence (ZAMS) abundances to PN abundances are
expected. If the progenitors had a typical halo or thick disc
compositon the abundance ratios would be similar to those found in
disk PNe (He/H$\sim0.100$, C/O$<1$ and N/O$\sim0.4$). The origin of
these PNe requires an explanation.

\citet{jac} surveyed 130 GCs in a search for PN in GC. They concluded
that the number known is probably complete. They found this is less
than the expected number ($\approx16$) given the total luminosity of
the Galactic globular cluster system. This number is based on stellar
populations which are younger than the stars in globular clusters. GCs
have very low mass turn-offs ($\sim0.85\,{\rm M}_{\sun}$). At this
mass a typical PN central star (CSPN) should have a mass of
$\sim0.52\,{\rm M}_{\sun}$. The time required from when a star of this
  mass leaves the asymptotic giant branch (AGB) for this star to get
  hot enough to ionize the ejected mass would be too long. The ejected
  mass would dissipate before a visible PN could be observed. This is
  known as a ``lazy'' planetary nebula \citep{ren79}.

The masses of the central stars of the globular cluster PN are larger
than the typical measured masses of the youngest white dwarfs (WDs) at
the top of the cooling sequence. The cooling sequences of white dwarfs
in globular clusters have been determined by a number of authors
\citep{renwd,cool,richwd,zocwd,han02,han04,han07,calwd}. They find
that the majority of main sequence stars evolve into white dwarfs of
mass 0.5-0.55$\,{\rm M}_{\sun}$. and the average of a young GC white
dwarf is probably between 0.50 and 0.53$\,{\rm M}_{\sun}$. All of the
GC PNe have central star (CSPN) masses above this
average. \citet{al} and \citet{bitwo} find that the mass of
the central star of Ps 1 is 0.58-0.60 M$_{\sun}$. \citep{jac}
determined the CSPN of JaFu 1 and JaFu 2 have masses of 0.55
M$_{\sun}$. The estimated mass of GJJC-1 is around 0.56 M$_{\sun}$
\citep{pen92}.

Both the CSPN masses higher than the white dwarfs and the unusual
abundances seem to require unusual stars as progenitors. GCs contain a
variety of unusual star types in addition to the standard types of
stars which should be considered as potential progenitors of GC
PNe. GCs contain blue stragglers which are thought to be the mergers
of two main sequence stars and hence act like main sequence stars with
masses higher then the turn-off mass. \citet{bitwo} and \citet{al}
suggested the projenitor of Ps 1 is a blue straggler because of its
high core mass and evidence of at least one third dredge up (TDU)
event as evidenced by the very high C/O ratio in the
nebula. Theoretical models of thermally pulsing-AGB (TP-AGB) stars
suggest a core mass of $\sim0.58\,{\rm M}_{\sun}$ is required to get a
TDU event.

Another type of star found in globular clusters are second generation
stars which incorporate material from the more massive stars of the
first generation (primordial component or P). The first generation
stars have abundances which reflect the abundances of the interstellar
medium material from which they formed and hence have a normal amount
of helium ($Y\approx0.25$). The second generation often
incorporates material with a higher fraction of helium than normal
($Y\approx0.30$). Some clusters show evidence for additional
populations with even higher $Y$ values (E.g. \citet{cal07}). These
multiple populations show up observationally in a number of ways. It
shows up as an Na-O anticorrelation in both red giants and in main
sequence stars. This was first observed by \citet{grat} when they
noted the Na-O anticorrelation shows up in main-sequence stars in
addition to red giants in several clusters. Some clusters have
distinct multiple main sequences (e.g. $\omega${Cen} \citep{bed04},
NGC 2808 \citep{pio07}, etc.)  which can be fit by having a second
main sequence with a higher $Y$. There are GC with multiple subgiant
branches (e.g. M22 \citep{mil10a,mil10b}, NGC 104, etc.). See
\citet{sgrev09} for a review of the evidence.

These second generation stars form a substantial portion of all stars
in GC (up to 60-70 percent of the total number of stars in a GC
\citep{caret}) and should have a substantial impact on what is
observed during the AGB and PN phases of evolution. In this paper I
look at the expectations of the PN phase from all the generations of
GC stars. I also model the expected PN phase of the blue straggler
stars. In section 2 I describe the TP-AGB model used. In section 3 I
describe the results of these models. In section 4 I discuss some
implications. In section 5 I summarize the results.

\section{Models}

Most of the relevant details of this model are explained in
\citet{bu97}, \citet{bue97}, and \citet{gbm}. In this section I
concentrate on modifications of this model. Of particular importance
is the mass-loss on the RGB and E-AGB which can be significant input,
especially at the low ZAMS masses of globular cluster
stars. Particular attention is paid to the effect of enhanced helium
abundances on mass-loss during these stages. The mass-loss in these
stages is found by integrating the mass-loss rate formula over the
Padua tracks. There is an important point to note, in the pre-TP-AGB
mass-loss model the stellar evolution and the mass-loss are not
coupled and hence the equations for mass-loss derived below should be
used with care. The mass-loss shift is probably a small effect as
explained below but the reader should be aware of it.

\subsection{Red giant mass-loss}
The mass-loss which occurs on the red giant branch (RGB) is very
important for low mass stars found in globular clusters, in some
extreme cases it may sometimes prevent the star from even reaching the
TP-AGB. Most of the pre-TP-AGB mass-loss occurs during the RGB. The
standard method to determine the amount of mass-loss is to use
Reimers' Law \citep{rei} given by
\begin{equation}
\dot{M}=\eta\frac{LR}{M}
\end{equation}
where L, R and M are the stellar luminosity, radius and mass,
respectively in solar units. However, to calculate the pre-TP-AGB
mass-loss the modified version of the Reimers formula of \citealt{sch}
given by:
\begin{equation}
\dot{M}=\eta\frac{LR}{M}(\frac{T_{\rm eff}}{4000{\rm K}})^{3.5}
(1+\frac{g_{\sun}}{4300g_{\star}})
\end{equation}
where $T_{\rm eff}$ is the effective stellar temperature, $g_{\star}$ is
the surface gravity of the star in cgs units. Values of 27400${\rm
  cms^{-2}}$ for $g_{\sun}$ and $8.0\times10^{-14}$ for $\eta$ were
adopted (Value recommended by \citealt{sch}). This new mass-loss rule
appears to give better results for horizonatal branch masses then the
Reimer's rate \citep{sch}.

This mass-loss law is applied to the variable $Y$ stellar evolution
tracks from the Padova stellar evolutionary library
(http://pleadi.pd.astro.it) described in detail in \citet{berta} and
\citet{bertb}. To determine the red giant mass-loss the mass-loss rate
was integrated from the beginning of the red giant branch (encoded in
the Padova files as brgbs) up to the tip of the red giant branch
(encoded as trgb) using the trapezoidal rule. The amount of mass-loss
between time steps in the models is given by:
\begin{equation}
\Delta M_{i}=\frac{1}{2}(\dot{m}_{i+1}+\dot{m}_i)(t_{i+1}-t_{i})
\end{equation}
where $t_i$ and $t_{i+1}$ are the model times and $\dot{m}_{i+1}$ and
$\dot{m}_i$ are the mass-loss rates at the corresponding times. The
total mass-loss is determined by summing all of the $\Delta M_{i}$s.

Table~\ref{padovamodels} shows the values of $Y$ and $Z$ for which
mass-loss was computed. For all available masses the mass-loss on both
the EAGB and RGB was computed. $Z$ is the value of $Z$ on the ZAMS and
$Y$ is the value of $Y$ on the ZAMS.

\setcounter{table}{0}
\begin{table*}
\centering
\caption{Padova Models Used}
\label{padovamodels}
\begin{tabular}{cc}
$Z$&$Y$\\\hline
0.0001&0.23,0.26,0.30,0.40\\
0.0004&0.23,0.26,0.30,0.40\\
0.001&0.23,0.26,0.30,0.40\\
0.004&0.23,0.26,0.30,0.40\\
0.008&0.23,0.26,0.30,0.40\\
0.017&0.23,0.26,0.30,0.34,0.40\\
\end{tabular}
\end{table*}
\medskip

\begin{figure*}
\includegraphics[width=188mm]{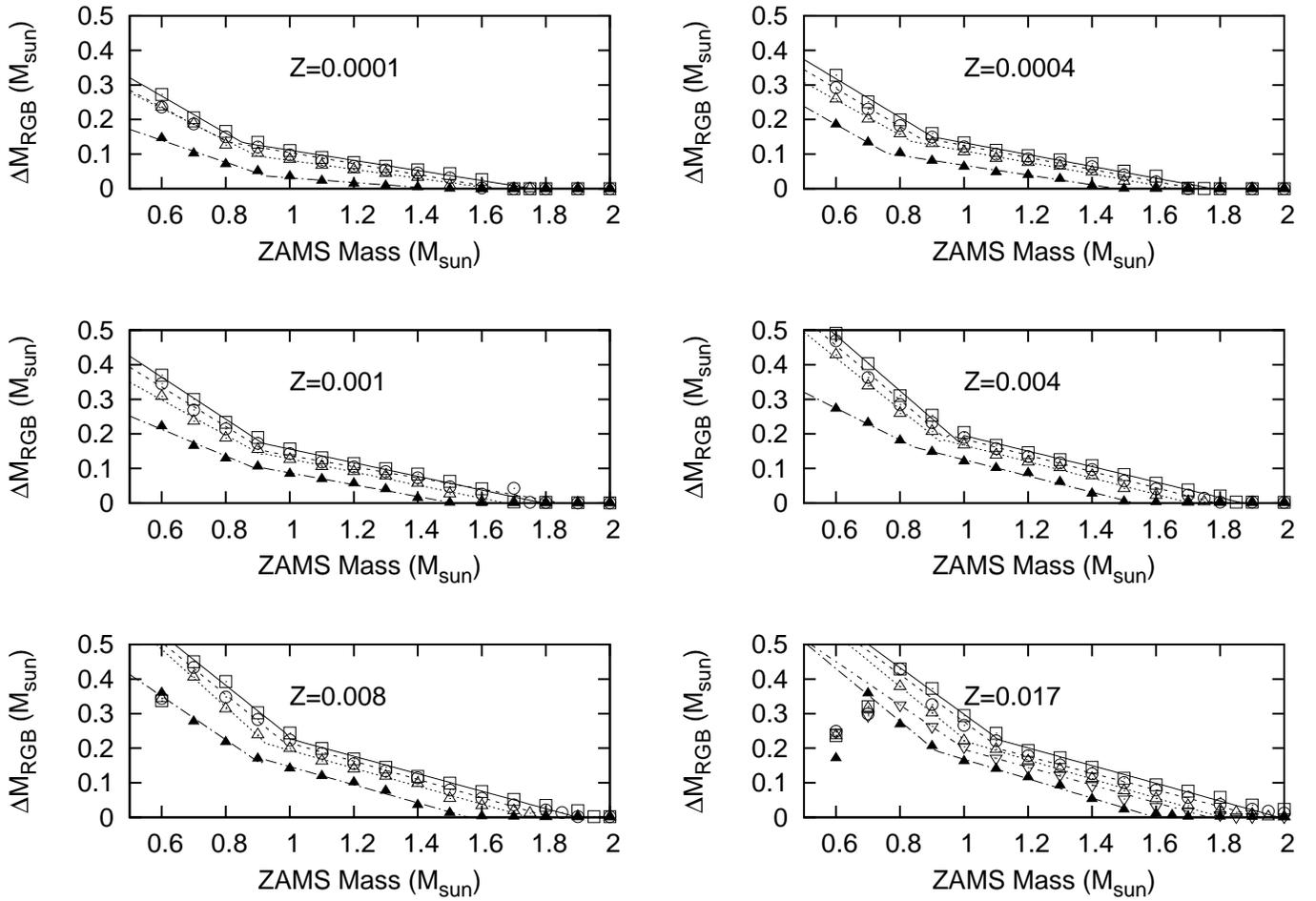}
\caption{Each panel in the figure shows the calculated mass-loss on
  the red giant branch for the $Z$=0.0001, 0.0004, 0.001, 0.004, 0.008
  and 0.017 models for different available values of $Y$. The open
  squares, open circles, open triangles and closed squares are the
  mass-losses for $Y$=0.23, 0.26, 0.30 and 0.40, respectively. The
  upside down triangles in the $Z=0.017$ panel are the mass-losses for
  the $Y=0.34$ models. The solid, dashed, dotted, and dot long dashed
  lines are the fits for the $Y$=0.23, 0.26, 0.30, and 0.40 mass-losses,
  respectively. The dot dashed line is the fit for the $Y$=0.34 models
  in the $Z$=0.017 panel.}
\label{rgbmassloss}
\end{figure*}

The mass-loss on the RGB as a function of the ZAMS mass for all
available values of $Y$ and $Z$ is shown in
figure~\ref{rgbmassloss}. In all panels it is evident the amount of
mass-loss decreases as $Y$ increases. This occurs because stars with
higher values of $Y$ means the RGB star will have smaller radii and
higher surface gravity due to the lower opacity in the outer
layers. These factors lower the mass-loss rates and the total
mass-loss.

The RGB mass-loss was fit using two linear fits for higher and lower
ZAMS masses. The transition point between the fits was determined by
visually estimating the mass where the slope appears to change. This
mass is typically found around a ZAMS mass of 0.8-0.9 M$_{\sun}$. The
higher mass fit was terminated where the high mass line crosses the
horizontal axis. This termination point was also estimated
visually. For masses larger the mass where the higher mass fit crosses
the axis the mass-loss is 0. The equations of the linear fits for low
and high masses are given by $\Delta{\rm M}_{\rm RGB,low}$ and
$\Delta{\rm M}_{\rm RGB,high}$. To make the fits work at the two
highest metallicities ($Z=0.008$ and 0.017) the fitting was done by
excluding the ${\rm M}=0.6\,{\rm M}_{\sun}$ models. These can be
safely excluded since the mass-loss for these very low mass stars
eliminates their envelope before the tip of the RGB is reached and
such models will not be considered in this paper. Visual inspection
indicates the fits are in good agreement to the mass-loss
calculations.

The equations for $\Delta{\rm M}_{\rm RGB,low}$ and $\Delta{\rm
  M}_{\rm RGB,high}$ are given by
\begin{equation}
\Delta{\rm M}_{\rm RGB,low}=a_{11}M+a_{10}
\end{equation}
\begin{equation}
\Delta{\rm M}_{\rm RGB,high}=a_{21}M+a_{20}
\end{equation}
The mass-loss is found by calculating the value of both $\Delta{\rm
  M}$s and finding the maximum value. If the mass-loss is found to be
negative then the value of the mass-loss is set to 0.  The
coefficients of these equations for the different values of $Y$ and
$Z$ are shown in table~\ref{tab:masslosscoes}.

\begin{table*}
\caption{Red giant mass-loss coefficients}
\label{tab:masslosscoes}
\begin{tabular}{cccccc}
$Y$&$Z$&$a_{11}$&$a_{10}$&$a_{21}$&$a_{20}$\\\hline
0.0001&0.23&-0.53375&0.587959&-0.146583&0.257894\\
0.0001&0.26&-0.48794&0.528645&-0.158946&0.263457\\
0.0001&0.3&-0.461671&0.509082&-0.128568&0.210224\\
0.0001&0.4&-0.317215&0.330223&-0.0682917&0.0996025\\
0.0004&0.23&-0.556562&0.651869&-0.171144&0.304073\\
0.0004&0.26&-0.54755&0.618164&-0.173845&0.298794\\
0.0004&0.3&-0.50535&0.560052&-0.166315&0.275781\\
0.0004&0.4&-0.52254&0.499175&-0.142323&0.210173\\
0.001&0.23&-0.607168&0.728346&-0.197265&0.353301\\
0.001&0.26&-0.571589&0.678889&-0.161241&0.298386\\
0.001&0.3&-0.509972&0.604365&-0.190175&0.317975\\
0.001&0.4&-0.384763&0.443838&-0.170916&0.258054\\
0.004&0.23&-0.56752&0.764646&-0.223724&0.418548\\
0.004&0.26&-0.65305&0.815914&-0.234945&0.419982\\
0.004&0.3&-0.658095&0.794849&-0.236612&0.403639\\
0.004&0.4&-0.45996&0.550102&-0.234695&0.359844\\
0.008&0.23&-0.73907&0.973301&-0.252334&0.479442\\
0.008&0.26&-0.75636&0.960051&-0.263265&0.478337\\
0.008&0.3&-0.831625&0.984962&-0.259694&0.453806\\
0.008&0.4&-0.629659&0.728283&-0.262052&0.40875\\
0.017&0.23&-0.673745&0.971984&-0.255924&0.50791\\
0.017&0.26&-0.682876&0.959522&-0.241278&0.465487\\
0.017&0.3&-0.75486&0.982077&-0.27485&0.495803\\
0.017&0.4&-0.7626&0.887937&-0.280617&0.448454\\
0.017&0.34&-0.6274&0.826845&-0.256405&0.452657\\
\end{tabular}
\end{table*}

No attempt has been made yet to calibrate this mass-loss, which will
be done in a later paper. However, the mass-loss values from these
equations appear to be reasonable. For example a $1.0\,{\rm M}_{\sun}$
$Y=0.26$ $Z=0.017$ star would experience 0.28$\,{\rm M}_{\sun}$ of
mass-loss on the RGB which is typical of other models. A typical
turn-off mass of $0.80\,{\rm M}_{\sun}$ with $Y=0.245$ and $Z=0.0008$
gives a RGB mass-loss of 0.22$\,{\rm M}_{\sun}$ which is reasonable
giving a zero-age horizontal branch mass of approximately 0.58$\,{\rm
  M}_{\sun}$ which is similar to measured values
(e.g. \citet{grat10}).

It should be noted, as suggested by the referee, that the method used
to find the mass-loss is not consistent with the stellar evolution
models. As the star loses mass its surface gravity would decrease
causing the star to expand. For the models used this would result in a
higher mass-loss rate near the tip of the RGB and a greater amount of
mass-loss on the RGB (and the E-AGB) then is calculated here. However,
this effect should be relatively small since the deviation will only
be really significant at the tip of the RGB. Although the method used
here is not strictly consistent the relative differences in mass-loss
due to the effect of the ZAMS helium abundances and the ZAMS
metallicity should be correct.

\subsection{First dredge-up}

This is important since when all stars enter the RGB the convective
envelope penetrates into regions of the star where partial H burning
has occurred bringing these products up to the surface. If a PN is
formed this process will have modified the surface abundances. This
paper only considers the effects on He and the CNO elements since
these are what is observed in planetary nebula. The first dredge-up
(FDU) prescription of \citet{gj94} is used.

\subsection{Mass-loss on the early-AGB}

The same procedure was applied to the early-AGB (E-AGB) portions of the
Padova tracks. An additional condition of starting the mass-loss when
the temperature was below 4500K was assumed since this mass-loss law
is applicable only to K and M stars.

Figure~\ref{eagbmassloss} shows the calculated mass-loss during the
E-AGB and the fits to these mass-losses. The most obvious trend is as
$Y$ increases so does the amount of mass-loss. This is opposite to the
trend on the RGB. In this case the core mass on the AGB is increased
which also increases the luminosity. This increase in the luminosity
on the E-AGB results in greater mass-loss. This enhancement of
mass-loss on the E-AGB is important since it means the higher the
value of $Y$ the more mass is lost on the E-AGB. It means such a star
has a higher probability to reach the horizontal branch but its
envelope may not survive to reach the first thermal pulse.

The mass-loss on the E-AGB is fit using 4 fits in different regions of
mass. The lowest mass range (${\rm M}\la1.5\,{\rm M}_{\sun}$) is fit
via a cubic, the next mass range up ($1.5\,{\rm M}_{\sun}\la{\rm
  M}\la2.0\,{\rm M}_{\sun}$) is fit using a quadratic fit. The next
mass range up ($2.0\,{\rm M}_{\sun}\la{\rm M}\la4.5\,{\rm M}_{\sun}$)
is fit using a linear fit. Finally the highest masses are fit using a
constant value of mass-loss. The points of intersection between
adjacent fits were visually estimated. This procedure gives a good fit
to the model mass-losses.

The equations for the E-AGB mass-loss in the first two mass regions
are given by:
\begin{equation}
\Delta{\rm M}_{\rm E-AGB}=a_{13}M^3+a_{12}M^2+a_{11}M+a_{10}
\end{equation}
and
\begin{equation}
\Delta{\rm M}_{\rm E-AGB}=a_{22}M^2+a_{21}M+a_{20}
\end{equation}
where $M$ is the mass of the star on the ZAMS. Only the coefficients
of first two regions have been included in table~\ref{tab:eagbco} to
save space and since no models of sufficient mass which need the fits
for the upper regions are calculated in this paper. The E-AGB
mass-loss is calculated by finding the intersection of the two regions
and then choosing the appropriate region and plugging into the
corresponding equation.

\begin{figure*}
\centering
\includegraphics[width=188mm]{eagbml.eps}
\caption{Each panel in the figure shows the calculated mass-loss on
  the early asymptotic giant branch as a function of ZAMS mass for the
  $Z=$0.0001, 0.0004, 0.001, 0.004, 0.008 and 0.017 models for different
  available values of $Y$. The symbols and lines have the same meaning
  as those in figure 1.}
\label{eagbmassloss}
\end{figure*}

\begin{table*}
\caption{Coefficients for fits to early-AGB mass-loss}
\label{tab:eagbco}
\begin{tabular}{ccccccccc}
$Z$&$Y$&$a_{13}$&$a_{12}$&$a_{11}$&$a_{10}$&$a_{22}$&$a_{21}$&$a_{20}$\\\hline
0.0001&0.23&-0.0217193&0.0915006&-0.122712&0.0744702&0.130361&-0.514963&0.508129\\
0.0001&0.26&-0.215429&0.683192&-0.699251&0.256586&0.0581723&-0.180144&0.136288\\
0.0001&0.3&-0.00119991&-0.0408501&0.10212&-0.0248815&0.014027&0.0478973&-0.123886\\
0.0001&0.4&-0.0758491&0.117157&0.0797828&-0.0451725&-0.4921&1.89761&-1.70715\\
0.0004&0.23&-0.0938861&0.306061&-0.327053&0.139884&0.111841&-0.424637&0.408071\\
0.0004&0.26&-0.0451582&0.138365&-0.144856&0.0819229&0.0275978&-0.0399283&-0.00915985\\
0.0004&0.3&-0.0753496&0.168929&-0.0967276&0.044191&0.0182938&0.033923&-0.0942867\\
0.0004&0.4&0.174151&-0.704201&0.939363&-0.300207&0.2393&-0.69827&0.619354\\
0.001&0.23&-0.0212184&0.0804842&-0.116847&0.0827674&0.0494489&-0.156333&0.123733\\
0.001&0.26&-0.0886186&0.296724&-0.336025&0.155924&0.0433492&-0.110378&0.0630566\\
0.001&0.3&-0.129515&0.334909&-0.262369&0.0984224&0.0160745&0.0320341&-0.0903071\\
0.001&0.4&0.288478&-1.1637&1.43381&-0.425362&-0.00537448&0.133165&-0.102088\\
0.004&0.23&0.00637805&-0.00833866&-0.0338022&0.0608243&0.0422666&-0.168079&0.171061\\
0.004&0.26&0.00369947&-0.00247321&-0.0397723&0.0679063&0.0672259&-0.27111&0.279355\\
0.004&0.3&-0.0537507&0.173439&-0.212193&0.129599&0.0521944&-0.184787&0.173164\\
0.004&0.4&-0.689675&1.65297&-1.08412&0.249005&0.0971864&-0.288754&0.274915\\
0.008&0.23&-0.0179112&0.0872929&-0.148306&0.102219&0.0335246&-0.1486&0.168721\\
0.008&0.26&-0.00338358&0.0399171&-0.111972&0.103433&0.0289011&-0.120242&0.132401\\
0.008&0.3&-0.0206433&0.0880253&-0.162452&0.134948&0.0418959&-0.164776&0.172505\\
0.008&0.4&1.46767&-5.9136&7.68436&-3.10794&0.109614&-0.405738&0.433205\\
0.008&0.34&-0.012937&-0.00120956&-0.0132868&0.0976855&0.0580864&-0.214707&0.216725\\
0.017&0.23&-0.00576822&0.041759&-0.106142&0.100898&0.0184395&-0.0922404&0.120053\\
0.017&0.26&-0.053432&0.248533&-0.399333&0.241192&0.0217703&-0.102463&0.128058\\
0.017&0.3&0.089118&-0.258045&0.170682&0.042502&0.0213919&-0.0934928&0.113367\\
0.017&0.4&-0.149333&0.461895&-0.471668&0.256413&0.0343459&-0.145229&0.203376\\
0.017&0.34&0.399763&-1.37573&1.41367&-0.348366&0.0352993&-0.143105&0.163442\\
\end{tabular}
\end{table*}

\subsection{Core mass at the first pulse}
From the Padova stellar evolution library I also extracted the core
mass as a function of the mass. Figure~\ref{coremass} shows the mass
of the carbon-oxygen core as a function of mass for several values of
$Y$ and $Z$. An important point to note is that as the initial helium
mass fraction increases so does the mass of the core. The core mass is
important since it is the most important factor controlling the
luminosity of an AGB star.

Figure~\ref{coremass} show the core mass at the first pulse from the
Padova models for $Z=$0.0001, 0.0004, 0.001, 0.004, 0.008, and 0.017
and $Y=$0.23, 0.26, 0.30 and 0.40 as a function of mass. Each set of
models with a given $Y$ and $Z$ have been fit by a double quadratic
fit, one at lower masses $\la1.5\,{\rm M}_{\sun}$ and one at the
higher masses.  The transition between the two was found between 1.3
and 2.0 ${\rm M}_{\sun}$. The transition between the lower mass and
higher masses was determined by visual inspection of where the core
mass begins to rise steeply. In all cases the fits to the points are
good giving core masses which are typically less than 0.02${\rm
  M}_{\sun}$ difference.

The equations of the quadratic fits for low and high masses are given
by ${\rm M}_{\rm c0,low}$ and ${\rm M}_{\rm c0,high}$. The equations are:
\begin{equation}
{\rm M}_{\rm c0,low}=a_{12}M^2+a_{11}M+a_{10}
\end{equation}
\begin{equation}
{\rm M}_{\rm c0,high}=a_{22}M^2+a_{21}M+a_{20}
\end{equation}
where M is the ZAMS mass. The coefficients for the different values of
$Y$ and $Z$ are shown in table~\ref{tab:mczero}. The procedure used is to
find the point of intersection between the two fits and then to plug
in the relevant mass.

\begin{figure*}
\includegraphics[width=168mm]{mczero.eps}
\caption{The figure shows the core mass at the onset of the first
  pulse for the $Z=$0.0001, 0.0004, 0.001, 0.004, 0.008, and 0.017
  models. The symbols have the same meaning as they do in
  figure 1.}
\label{coremass}
\end{figure*}

\begin{table*}
\caption{Coefficients for the fits to the first pulse core masses}
\label{tab:mczero}
\begin{tabular}{cccccccc}
$Y$&$Z$&$a_{12}$&$a_{11}$&$a_{10}$&$a_{22}$&$a_{21}$&$a_{20}$\\\hline
0.0001&0.23&-0.0267946&0.123057&0.442823&-0.0241626&0.283874&0.147144\\
0.0001&0.26&-0.0449264&0.172855&0.417791&-0.0340353&0.333449&0.114275\\
0.0001&0.3&-0.0317268&0.175202&0.407754&-0.0236789&0.261573&0.255724\\
0.0001&0.4&-0.131429&0.4672&0.278697&-0.0276698&0.257098&0.381668\\
0.0004&0.23&-0.0596658&0.182913&0.415731&-0.0279409&0.313233&0.0872891\\
0.0004&0.26&-0.0775454&0.235014&0.391988&-0.0373996&0.359006&0.0649862\\
0.0004&0.3&-0.105043&0.31126&0.357675&-0.0265437&0.282672&0.216724\\
0.0004&0.4&-0.290637&0.721285&0.197312&-0.0194077&0.213603&0.429181\\
0.001&0.23&-0.0891579&0.224579&0.400277&-0.0313901&0.344189&0.0106343\\
0.001&0.26&-0.10291&0.267182&0.381744&-0.0282899&0.320649&0.0719709\\
0.001&0.3&-0.145894&0.366037&0.342087&-0.0365283&0.353432&0.0866877\\
0.001&0.4&-0.228835&0.580823&0.28018&-0.020842&0.226895&0.391846\\
0.004&0.23&-0.0830364&0.196981&0.413575&-0.0364602&0.399655&-0.166655\\
0.004&0.26&-0.0824008&0.201012&0.414887&-0.039857&0.415317&-0.159354\\
0.004&0.3&-0.117928&0.279869&0.387053&-0.0250966&0.306343&0.0555899\\
0.004&0.4&-0.139744&0.34628&0.413321&-0.0323089&0.3129&0.197688\\
0.008&0.23&-0.0624872&0.153486&0.428388&-0.0240687&0.314676&-0.071922\\
0.008&0.26&-0.0680031&0.165896&0.425421&-0.0289697&0.34486&-0.0954135\\
0.008&0.3&-0.0768498&0.187131&0.429621&-0.0256837&0.314658&-0.00601164\\
0.008&0.4&-0.118314&0.290786&0.436894&-0.0155313&0.224985&0.2672\\
0.017&0.23&-0.0403038&0.110723&0.443374&-0.0130873&0.233825&0.00316468\\
0.017&0.26&-0.0494165&0.129165&0.442886&-0.0186948&0.275689&-0.040745\\
0.017&0.3&-0.0561381&0.142527&0.441254&-0.0198798&0.266728&0.036109\\
0.017&0.4&-0.10803&0.266918&0.428987&-0.00752814&0.166398&0.315329\\
0.017&0.34&-0.0844639&0.200431&0.435484&-0.0139933&0.218876&0.161534\\
\end{tabular}
\end{table*}

The most obvious trend is there is an increase in the core mass as the
value of $Y$ increases. This is important since on the AGB a larger core
mass leads to a higher luminosity. This is also important since the
mass at the first pulse is an important factor in determining the mass
of the CSPN.

\subsection{Third dredge up}
In synthetic AGB models the standard method to model the third dredge
up (TDU) effect is to use a dredge-up parameter $\lambda$ so that
\begin{equation}
\lambda=\frac{\Delta M_{\rm dredge}}{\Delta M_{\rm c}}
\end{equation}
where $\Delta M_{\rm dredge}$ is the mass dredged up and $\Delta
M_{\rm c}$ is the increase in the core mass during the preceding
interpulse phase. During a thermal pulse the star develops a
convective shell in the region between the intershell region between
the base of the hydrogen-rich envelope and just above the core. This
region is helium and carbon rich since it consists of the products of
partial helium burning. At the end of the thermal pulse the convective
envelope may penetrate into this region and mix this carbon and helium
rich material into the envelope. The parameter $\lambda$ is a measure
of how deeply the convective envelope penetrates into this intershell
region and determines how much mass is mixed up into the outer layers.

A number of authors have used synthetic models to constrain the value
of $\lambda$ and the minimum core mass value at which TDU can occur,
M$_{\rm c,min}$ (e.g. \cite{gj93};\cite{mar99}). These authors used
the LMC and SMC carbon star luminosity functions to constrain both
${\rm M}_{\rm c,min}$ and $\lambda$ parameters. They show the value of
M$_{\rm c,min}$ is approximately $0.58\,{\rm M}_{\sun}$. In this paper
the value of ${\rm M}_{\rm c,min}$ is treated as a free parameter but
values near this are always chosen. The value of $\lambda$ is a free
parameter but it will always be small since most models will
experience only one TDU and I expect the value of $\lambda$ would grow
during subsequent thermal pulses if they were to occur.

\subsection{Mass-loss on the TP-AGB}

On the TP-AGB, mass-loss is calculated by the pulsation period-mass
loss law of \citet{vw93} without their correction for periods above
500 days. To make the transition from the modified Reimer's rate to
this pulsation mass-loss rule the modified Reimer's rate is used until
the pulsation mass-loss rule becomes larger.

\subsection{Conditions for formation of a visible planetary nebula}

The formation of a visible PN requires that the central star of the
planetary nebula (CSPN) reach a temperature of approximately 30000K on
its blueward journey from the AGB to the white dwarf cooling tracks
before the ejected envelope has sufficient time to disperse into the
interstellar medium. In a globular cluster the typical star at the
turn-off would produce a CSPN of mass approximately $0.52\,{\rm
  M}_{\sun}$. 

The criteria used in this paper assumes that if a planetary nebula
does not form before the ejected envelope expands to $0.5\,{\rm pc}$
it will not be visible. If an expansion rate of $15\,{\rm km}\cdot{\rm
  s}^{-1}$ is assumed for PN then the maximum transition time to form
a visible planetary nebula is given by approximately 25000
years. Since PN with expansion ages of 30000 years exist this
conservative value is adopted.

\begin{figure*}
\includegraphics{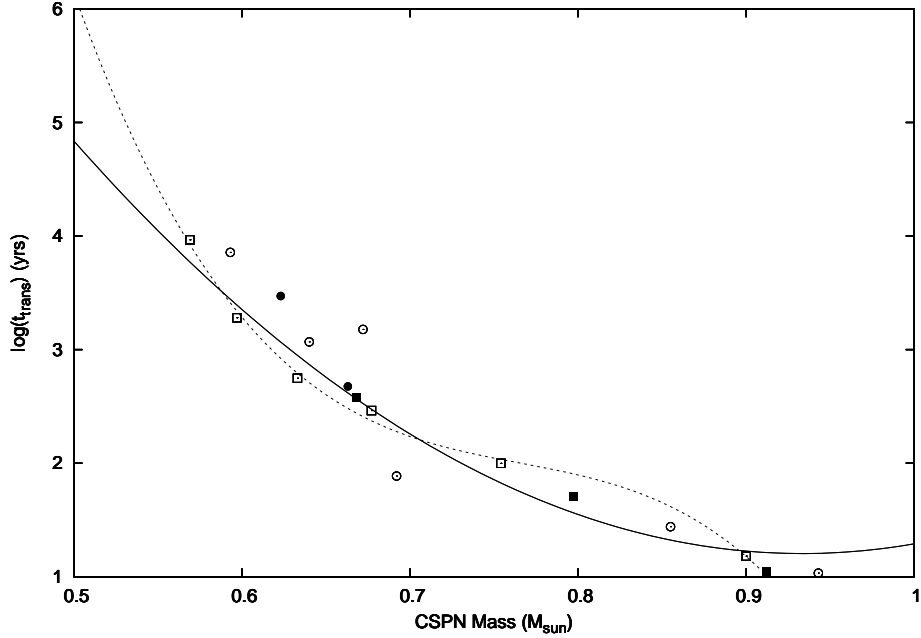}
\caption{The figure shows the transition times as a function of core
  mass. The open squares are the results calculated from the
  \citet{vw94} $Z=0.016$ models. The filled squares, open circles and
  filled circles are the results from the $Z=$0.008, 0.004 and 0.001
  models, respectively. The solid and dashed lines are the quadratic
  and cubic fits described in the text.}
\label{fig:transtime}
\end{figure*}

To find the approximate transition time for all the $Z=0.016$ models of
\citet{vw94} were linearly interpolated in time and $\log{{\rm
    T}_{\rm eff}}$ to find the time when the effective temperature becomes
30000K. This is the transition time $t_{\rm trans}$. In
figure~\ref{fig:transtime}, $\log{t_{\rm trans}}$ is plotted as a function
of the CSPN mass. The results for the $Z=0.016$ models are fitted with
both a quadratic and cubic fit. These fits are used to extrapolate an
approximate range of values for CSPN masses lower than 0.57$\,{\rm
  M}_{\sun}$ cores. Also included were the results from other
metallicities. With the exception of one point from the $Z=0.004$
models (their ${\rm M}=2.0\,{\rm M}_{\sun}$ model) all of the points
follow the trend of the cubic.

The quadratic fit is given by:\\
\begin{equation}
\log{t_{\rm trans}}=19.3316M_{\rm c}^2-36.0874M_{\rm c}+18.0464
\end{equation}
The cubic fit is given by:\\
\begin{equation}
log{t_{\rm trans}}=-181.893M_{\rm c}^3+417.538M_{\rm
  c}^2-322.289M_{\rm c}+85.6
\end{equation}
The cubic fit is a better fit in that it passes closer to all the
points than does the quadratic fit. The cubic fit seems to better
capture the extrapolation, however, since these are used for
extrapolation caution needs to be exercised and the values derived
here should be treated as approximate. If the transition time is set
to 30000 years and the quadratic and cubic equations are solved for
$M_c$ the results are 0.522${\rm M}_{\sun}$ and 0.548${\rm M}_{\sun}$,
respectively. The cubic extrapolation is weighted more heavily and the
adopted minimum value of the CSPN mass to produce a visible PN is
0.545${\rm M}_{\sun}$. This value is very close to the lowest observed
value of a globular cluster CSPN central masses, JaFu 1 and JaFu 2. It
should be noted a choosing a different set of post-AGB models might
lead to different conclusion about the minimum CSPN mass because the
transition time in post-AGB models depends strongly on the adapted
mass-loss rate.

\subsection{TP-AGB models}

The TP-AGB is followed using a synthetic AGB code which is a
descendent of the \citet{rv81} code. The code begins with a guess at
${\rm T}_{\rm eff}$ to calculate the surface boundary condition. The
equations of stellar structure is then integrated to the base of the
convective envelope. The value of the effective temperature is
modified until the base of the convective envelope is at the same
position as the core mass. The opacities used for high temperatures
are the \citet{opal} and for low temperatures the opacities of
\citet{af94} are used. The luminosity of the star is calculated using
the expressions in \citet{wg98}. A mixing length parameter,
$\alpha=l/{\rm H}_p$, of 1.70 is used. This value is chosen since it
is close to typical values of values of $\alpha$ chosen for solar
models.

\section{Results}

\subsection{First Generation Star Models}
\begin{figure*}
\includegraphics{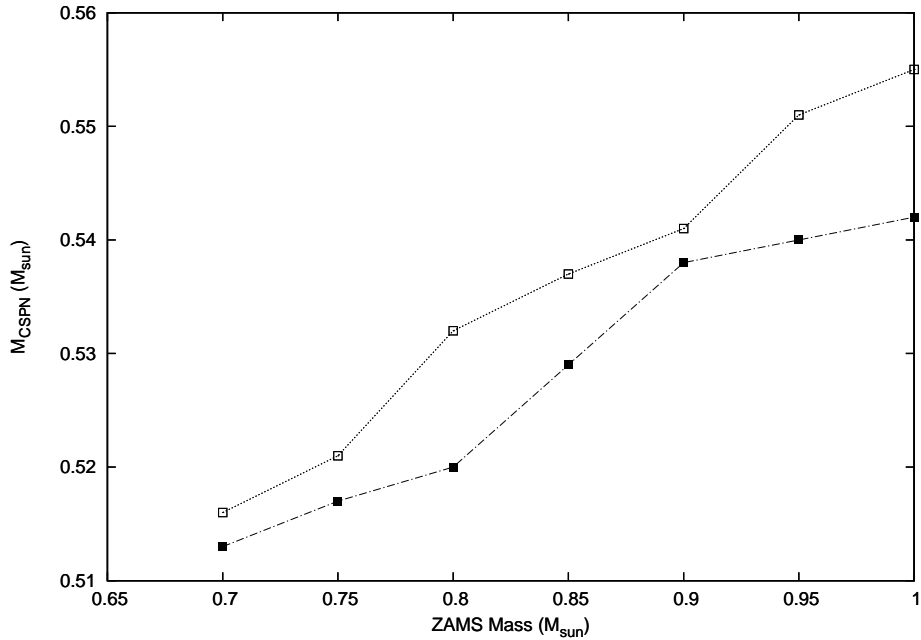}
\caption{This figure shows the model CSPN masses as a function of ZAMS
  mass. The open squares are the $Z=0.0002$ models and the closed
  squares are the $Z=0.004$ models.}
\label{fig:cspnzvar}
\end{figure*}

First generation stars are stars with the lowest possible $Y$ of a
globular cluster. A series of models with masses between 0.7 and 1.0
M$_{\sun}$ were calculated for metallicity values of 0.0002 and 0.004
and a typical $Y$ value of 0.245. Figure~\ref{fig:cspnzvar} shows the
final white dwarf mass as a function of the ZAMS mass for a variety of
$Z$ values and the $Y$ value of 0.23. If we adopt a typical globular
cluster age of $\sim13\,{\rm Gyr}$ it gives typical turn-off masses of
$\sim$0.85 M$_{\sun}$. Note for all values of the metallicity, $Z$,
the final mass is below the mass needed to produce a visible PN.

If these stars produced visible PNe then there would be a number of
PNe in globular clusters with He/H around 0.100 and values of N/O and
C/O similar to the values of a typical disk PN. Since the CSPN masses
are too low, these models suggest first generation stars do not
produce visible PNe as the ejected mass would disperse before it could
occur. This reduces the number of expected PNe in GCs since these
stars are too low mass to produce them. The larger the fraction of
primordial stars is the lower the number of expected PNe. Since no
typical PN are observed in the GC system this model explains this
observation.

The typical white dwarf mass which would result from these first
generation stars is about 0.525 M$_{\sun}$. These values are in rough
agreement with the typical values inferred for the WD mass at the top
of the cooling sequence determined from measurements. \citet{han07}
finds WD masses at the top of the WD cooling sequence of globular
cluster NGC 6397 between 0.50 and 0.53 M$_{\sun}$ although their error
analysis favors a lower value. \citet{han04} found the mass of the WD
at the top of the cooling sequence of globular cluster M4 as 0.55
M$_{\sun}$. \citet{str09} find the mass of the WD cooling sequence of
CO WDs in NGC 6397 is 0.53 M$_{\sun}$.

\subsection{Planetary Nebulae from Second Generation Stars}
\begin{figure*}
\includegraphics{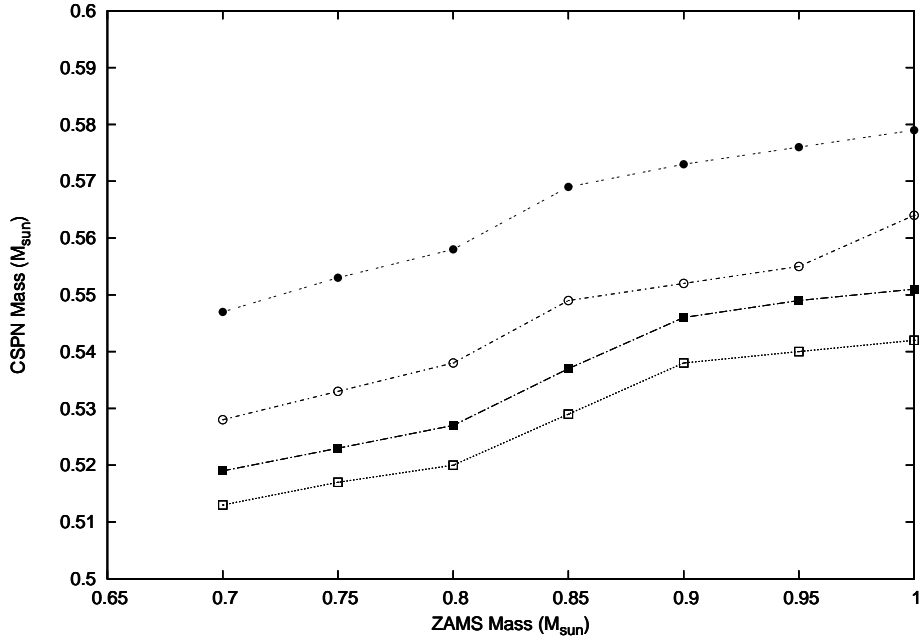}
\caption{Figure shows the model CSPN masses as a function of ZAMS mass
  for models with $Z=0.004$. The open squares, filled squares, open
  circles, and filled circles are the $Y=$0.244, 0.274, 0.304, and
  0.334 models, respectively.}
\label{fig:cmhe}
\end{figure*}

These stars start on the ZAMS with a high initial He abundance
($Y\approx0.30$). Figure~\ref{fig:cmhe} shows the calculated
initial-final relationships for ZAMS stars with masses ranging from
0.7 to 1.0 M$_{\sun}$ with $Z=0.004$ and a range of possible values of
$Y$. It shows as the initial helium abundance is increased the final
white dwarf mass increases. This effect is somewhat counter balanced
by the shorter ZAMS lifetimes of stars and lower masses of the
turn-off with higher initial $Y$. However even for an assumed turn-off
of 0.75 M$_{\sun}$ with the higher values of $Y$ the final mass is near
or above the limit to produce a visible PN.

For the stars with a primordial value of helium ($Y=0.245$), $Z=0.004$
and a turn-off mass of 0.85$\,{\rm M}_{\sun}$ the core mass here is
0.529$\,{\rm M}_{\sun}$ which would be too small to produce a visible
PN. These models suggest the reason only PN with high He/H appear is
this allows larger core masses; which lead to shorter transition times
allowing a visible PN to appear.

Since these He enhanced stars can be a significant fraction of the
total number of stars in a globular cluster it also predicts that
there where two separate populations with different values of $Y$ exist
there should should be two CO WD cooling tracks at slightly different
masses. Since it appears the typical WD mass is around 0.53$\,{\rm
  M}_{\sun}$ and 0.54-0.55$\,{\rm M}_{\sun}$ we estimate that these
tracks are separated by 0.01-0.02$\,{\rm M}_{\sun}$. This would be
observationally challenging to do but might be possible.

\begin{figure*}
\includegraphics{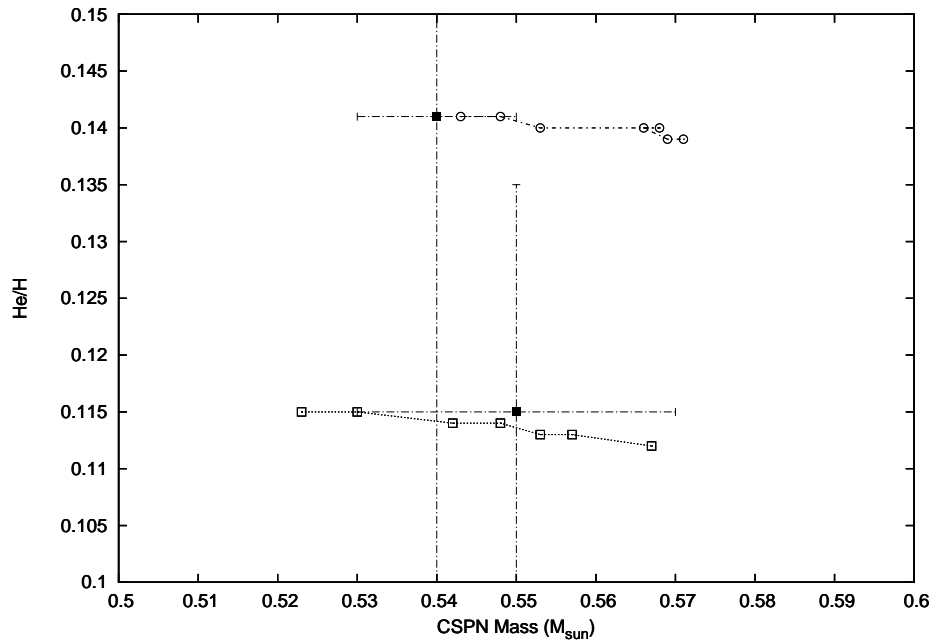}
\caption{This figure shows He/H as a function of CSPN mass for JaFu 1
  and JaFu 2 as filled squares. The models of 'best' fit for JaFu 1
  and JaFu 2 discussed in the text are shown as open circles and open
  squares, respectively. Each of the points on the model curves starts
  at 0.70$\,{\rm M}_{\sun}$ and is iterated by 0.05$\,{\rm M}_{\sun}$
  up to 1.00$\,{\rm M}_{\sun}$.}
\label{fig:pnfit}
\end{figure*}

Figure~\ref{fig:pnfit} shows the model grids used to fit the CSPN
masses and He/H for the model parameters which are the closest fit to
observed values of $M_{\rm CSPN}$, He/H and $\log{\rm O/H}+12$ from
Jacoby et al. of JaFu 1 and JaFu 2. For both nebula the ZAMS values of
$Z$ was varied until the model value of O/H was a close match to the
observed abundances of O/H found in the nebula (from Jacoby et
al.). Then the value of $Y$ was adjusted until a close fit to He/H was
obtained. The goal here was to find a reasonable set of parameters and
not to fine tune this to the best possible fit.

For JaFu 2 the best fitting model has a core mass of 0.55 M$_{\sun}$
which is the same as the observed value. This core mass is produced by
a model ZAMS star of mass approximately 0.85$\,{\rm M}_{\sun}$ which
is a quite reasonable turn-off mass. The values of the model ZAMS $Y$
and $Z$ are 0.282 and 0.00102, respectively. The value of $Y$ is quite
reasonable in this case being in line with which are quite reasonably
in line with expectations for helium enhancements. JaFu 2 is found in
globular cluster NGC 6441 which has an [Fe/H] of -0.46 (\citet{har96}),
which is a bit higher than the adopted value of $Z$ for the
nebula. However, [Fe/H] is not actually fit here and O/H is. It is
quite possible there are oxygen depletions which correspond to the
helium enhancement here over the different populations in this
cluster. This would be similar to what was found by \citet{pio07} for
NGC 2808. It is expected He is enhanced by the operation of the CNO
cycle which would result in a depletion of oxygen.

For JaFu 1 the best fitting model has a final CSPN mass of 0.54
M$_{\sun}$, $Y_{\rm ZAMS}=0.333$, $Z_{ZAMS}=0.006947$. The value of $Y$ is
typical of the expected $Y$ enhancements. However the ZAMS mass of the
``best'' model is 0.70$\,{\rm M}_{\sun}$ which is too small but the
errors are large enough here to allow a range of possible ZAMS masses
up to 0.90$\,{\rm M}_{\sun}$. The ZAMS mass of 0.90$\,{\rm M}_{\sun}$
gives a core mass of 0.558$\,{\rm M}_{\sun}$ which is a little bigger
than the observed value from \citet{jac} ($0.54\pm0.2\,{\rm
  M}_{\sun}$) but it is well within the error bars.

It is difficult to compare the metallicity of the model for JaFu 1 to
that of the cluster Pal 6. A range of potential values are found in
the literature. On the low end are the works of \citet{lc02} and
\citet{zinn85} who find [Fe/H] of $-1.08\pm0.06$ and -0.74,
respectively. On the high end are the works of \citet{bi98} who find
$[Z/Z_{\sun}]=-0.09$ and \citet{ort95} and \citet{min95} who find
[Fe/H]=-0.4 and +0.22, respectively. The oxygen abundance of JaFu 1
suggests an overall metallicity between 1/3 and 1/2 solar which is
within the metallicity range on the high end of the metallicity
possibilities.

The observed nebular mass of JaFu 2 (of 0.04$\,{\rm M}_{\sun}$ from
Jacoby et al.) is a reasonable match to the models. Low ZAMS mass
models with values of $Y$ around 0.30 predict a nebular mass is just a
few times $0.01\,{\rm M}_{\sun}$. However, the observed nebular mass
of JaFu 1 is much too high (0.40$\,{\rm M}_{\sun}$) but Jacoby et
al. determined this using an assumed filling factor of 1 and note a
filling factor of 0.1 gives a much lower mass.

These models provide a good fits to the most important parameters; the
CSPN mass, the observed value of He/H, the turn-off masses and the
oxygen abundance of these PNe. These two nebula can be explained with
reasonable progenitors and slightly enhanced values of He/H.

\subsection{High C/O Nebula}
No star with a mass below the turn-off of any globular cluster will
produce a carbon-rich planetary nebula like Ps 1 since a core mass of
approximately 0.58 M$_{\sun}$ is required for the third dredge-up to
occur and since globular cluster stars at the turn-off will have core
masses well below this. {\it There is however a channel to get to
  these masses - blue stragglers are thought to be merged stars and as
  such are more massive than stars on the turn-off.} Since blue
stragglers in principle could have masses up to twice that of the
turn-off these are the most likely channel.

The model used here is the same as discussed in \citet{bu97} where
after a dredge-up occurs the envelope is so polluted with carbon it is
essentially immediately ejected. As discussed by \citet{bu97} this
single dredge-up heavily pollutes a low metallicity envelope and the
metallicity of the envelope jumps to near solar values. This results
in an expansion of the star and this resulting expansion causes the
star to immediately switch into a superwind phase. The remaining
envelope is eliminated in a few thousand years suggesting the carbon
star phase would be shorter than the PN phase. This explains the lack
of carbon stars in globular clusters.

The models calculated here for low metallicity ($Z\approx0.0002$)
confirm this basic scenario. If ${\rm M}_{\rm c,min}$ is set to
$0.57\,{\rm M}_{\sun}$ then the minimum ZAMS mass is $1.15\,{\rm
  M}_{\sun}$.  Figure~\ref{fig:k648mod} shows the TP-AGB evolution of
the important parameters of this model. On the last pulse the envelope
is heavily polluted and note the resulting change in the mass-loss
rate. The envelope in this case is ejected in a few thousand
years. The model produce a CSPN of the correct mass and the C/O ratio
of the model compares favorably to observations of the final ejecta.

\begin{figure*}
\includegraphics{m1-15z0002mod.eps}
\caption{This figure shows the TP-AGB evolution of a ZAMS star
  M=1.15$\,{\rm M}_{\sun}$, $Z=0.0002$ and $Y=0.2454$. The top panel
  shows the evolution of the ratio of carbon to oxygen. The lower
  panel shows the evolution of the mass-loss rate (MLR). The points
  are the values at the end of each interpulse cycle.}
\label{fig:k648mod}
\end{figure*}

Since M15 contains a number of blue stragglers (\cite{yan94}), which,
in this model are the progenitor star for a high C/O nebular like Ps 1,
M15 is a {\it reasonable site} for the formation of this PN.

The hardest observation to match is the low nebular mass of Ps 1. In
the blue straggler model the predicted nebular mass is $\sim0.2\,{\rm
  M}_{\sun}$ whereas the observed nebular mass is about 1/10th this
value. The observed smaller nebular mass suggests for the blue
straggler model to be a viable model their must have been additional
mass-loss, perhaps due to binary interaction. There seems to be no way
to reproduce this observation with a single star.

An alternative scenario is described by \citet{ots08} who argue stars
like K648 have evolved from a binary and its progenitor is a CEMP-s
star. The problem for this scenario is to match the CSPN mass of K648
the ZAMS mass of the star needs to be larger then the turn-off
mass. Otherwise the CSPN mass would be smaller.

Another complication to the model scenario in this paper is if the
initial metallicity of the star is increased the minimum mass to
produce a carbon star gets larger. This occurs because the mass of the
core at the first pulse gets smaller with smaller metallicity. This is
consistent with studies which infer the minimum mass to produce a
carbon star in the SMC, LMC and the Galaxy where as the metallicy goes
up so does the minium mass (See - references here (Marigo,
Groenewegen, etc.)). In this model a blue straggler in a higher
metallicity cluster ($Z=0.004$) would be less likely to produce a
carbon-rich nebula. This blue straggler would produce a visible PN
with an oxygen-rich composition.

Blue stragglers are rare but prominent members of GCs and cannot be
expected to produce many PNe. A simple estimate can be obtained in the
following manner; assume the average lifetime of a blue straggler is
$10^9$ years and the lifetime of a PN is 25000 years. If it is assumed
every blue straggler produces a PN then we should expect 1 PN for
every 40000 blue stragglers. If we assume the number of blue
stragglers per cluster is 100-200 and use 150 for the number of
globular clusters then the total number of blue stragglers in the GC
system should be between 15000-30000. This would give the number of PN
from this channel as 0.375-0.70 PNe. The assumed numbers are probably
optimistic since the lifetime of blue stragglers could be larger; the
lifetime of PN smaller and the 100-200 is the number of blue
stragglers for bigger clusters. The actual number of expected PN may
be even smaller. If it is assumed there are sufficient numbers to
produce only one in the entire GC system then it is possible Ps 1 is
the only example and an oxygen-rich PN is equally likely.

This model shows it is possible to produce carbon-rich PNe in globular
clusters but it is not possible for it to match the nebular mass of
Ps 1, which can only be done by invoking an ad hoc binary
explanation. This may be reasonable since this star probably formed in
a binary but at the moment it is untestable since there are no similar
PN in the GC system to compare it to. A similar PN with a well
constrained CSPN mass and nebular mass would tell us if K648 is
typical of its class. The only possible known analogs are the halo PN
such as BB-1 and H4-1 which have C/O$>$1 (see Howard et
al.). Unfortunately the distances to these nebula is poorly known and
it will be difficult to get precise information about the CSPN and
nebular masses for comparison purposes.

\subsection{What is the origin of the planetary nebula in M22?}
M22 contains a very interesting PN GJJC-1 which is nearly hydrogen
free \citep{gil} because no nebular hydrogen lines have been
detected. The nebula appears to be overabundant in O and Ne. The
central star of GJJC-1 has been observed \citep{hp93} and it is
suggested that the surface abundances are He/H=0.5, C is 6 times solar
and N is 14 times solar (\citet{hp93}, \citet{rau98}). 

There are only two other hydrogen free nebula out of approximately
1000-2000 total known in the Galaxy. Given the inferred percentage of
a PN being hydrogen free of 0.1-0.2 percent it seems unlikely that
this PN simply occurred by chance. One speculative possibility is the
existence in some clusters of a third generation of stars which would
have very extreme ($Y=0.40$) helium enhancement.

Recent work on the cluster by \citet{dacos} and \citet{mar09}
indicates M22 has at least two separate populations which vary in
their value of [Fe/H]. The average [Fe/H] is between -1.7 and -1.9 but
it ranges up to -1.4. The higher [Fe/H] population probably has a
higher value of $Y$. It has been suggested M22 is the remains of a
captured dwarf elliptical galaxy. As such it seems a reasonable
candidate for a population with extreme helium enhancement. M22 may
also have a more complicated star formation history with the formation
of different populations spaced out by significant amounts time.

A model with ${\rm M}=0.80\,{\rm M}_{\sun}$, $Y=0.40865$, and
$Z=0.000886$ was run. In this model the star reached the horizontal
branch but the entire envelope was lost by the end of the E-AGB. The
mass-loss on the E-AGB increases as the value of $Y$ goes up because the
core mass goes up. The luminosity of an AGB star will depend strongly
on the core mass and this increased luminosity drives the increased
mass-loss. This model produces a core mass of 0.60$\,{\rm
  M}_{\sun}$. If the mass is lowered to $0.75\,{\rm M}_{\sun}$ then
the model once again produces a HB star but all the remaining envelope
mass is lost on the E-AGB. This lower mass produces a $0.58\,{\rm
  M}_{\sun}$ remnant.

Why is this significant? This is close to the inferred mass of
GJJC-1's CSPN and production of a hydrogen free nebula requires
getting rid of the initial hydrogen before the PN phase. Getting rid
of all the envelope means if this star experiences a thermal pulse
then there would be little to no hydrogen ejected and the star would
eject helium rich material and it would look much like GJJC-1. This
suggestion should be regarded as speculative and the positive evidence
for it is thin but the idea seems to be possible and merits additional
study.

\subsection{How many of each type?}

Without a detailed population study which is beyond the scope of this
paper it is only possible to show the numbers work out approximately
correct. I start by assuming the \citet{jac} statement that given the
total luminosity of the GC system then the number of PNe should be 16.
If it is assumed that 70 percent of globular cluster stars are
primordial and have a turn-off in the 0.80-0.90$\,{\rm M}_{\sun}$
range then this reduces the number to about 4-5 since these stars
produce none. The number from this appears to be 3 which is in rough
agreement. Further, assuming all blue stragglers produce a visible PN,
we assume about 1 from this part which matches the designation of Ps 1
(K648) as being produced by this channel. The rest will be produced by
second generation stars (which may or may not all produce a visible
PN). Therefore the numbers are roughly consistent with the number of
known PNe.

\section{Discussion}

What do JaFu 1 and JaFu 2 tell us about second generation GC stars?
They confirm, independently of the use of colour magnitude
diagrams, stars exist in globular clusters with $Y=0.28-0.33$ and it
suggests these second generation stars are a significant fraction of
the number of GC stars. These two PNe could turn out to be very
important since they allow the direct observation of elements which
can not be observed directly using stellar spectroscopy.

For JaFu 1 $\log{\rm N/O}=-0.52$ (Jacoby et al.) which is slightly
higher than $\log{\rm N/O}_{\sun}=-0.88$ \citep{asp05}.  This ratio is
consistent with both a first dredge-up event and possibly a small
amount of nitrogen enrichment or oxygen depletion. For JaFu 1
$\log{\rm S/O}=-1.35$ which is consistent with the solar $\log{\rm
  S/O}=-1.50$. From these abundance ratios either all these have been
enriched by the same relative amount or the star has not been enriched
relative to a first generation star in Pal 6. The sulfer abundance of
JaFu 1 is given by $\log{\rm S/H}-\log{\rm S/H}_{\sun}=-0.55$. This is
consistent with a cluster with metallicity between 1/3 and 1/2 solar.
If the cluster's metallicity is on the lower end of its range then
this would indicate all elements have been enhanced in this cluster.

JaFu 2 may indicate a depletion of oxygen in NGC 6441. The value of
$\log{\rm O/H}-\log{\rm O/H}_{\sun}=-0.93$ and the value of
$\log{\rm Ar/H}-\log{\rm Ar/H}_{\sun}=-0.72$. Since this is a lower
metallicity cluster we would expect oxygen as an alpha element to be
enhanced however it appears to be less enhanced relative to
argon. This would be consistent with the enhanced material having been
processed by CNO cycling.

As GCs age they leak stars into the field via collisions and also by
tidal stripping. In fact it is estimated that 4-9\% of halo stars are
2nd generation globular cluster stars \citep{ves10}. If second
generation halo stars become halo stars they could produce PN similar
to JaFu 1 and JaFu 2 and may be GJJC-1. However, this is a small
fraction of the halo stars and it is quite possible no such star has
been observed.

Other places to look for similar PN might be the system of satellite
galaxies. The GCs $\omega$ Centauri, M54 and M22 are all possible
captured satellite galaxies and all of them have multiple populations
(see Piotto 2009). This suggests multiple populations might be common
in satellite galaxies and this model predicts there may be PNe in
these satellite galaxies similar to the GC PNe.

\section{Conclusions}

This paper presents a series of models for the expected TP-AGB stars
in globular clusters. The results of these models are compared to the
observed abundances of the globular cluster PNe and the measured
masses of WDs.
\begin{enumerate}
\item{These models suggest the typical unenriched ZAMS star at the
  turn-off will not produce a visible PN, which explains small number
  of PN in GC. The CSPN mass is too small to produce a visible PN
  during the transition from the AGB to the white dwarf phase.}
\item{The PNe JaFu 1 and JaFu 2 are consistent with being produced by
  second generation stars in the globular cluster. Both have He/H
  ratios which are consistent with $Y_{\rm ZAMS}\approx0.30$ which is
  consistent with results for second generation stars from fitting
  colour-magnitude diagrams to the cluster.}
\item{Due to its large core mass (0.56-0.58$\,{\rm M}_{\sun}$) it is
  quite possible the PN GJJC-1 is consistent with a third generation
  ($Y\approx0.40$) progenitor.}
\item{The nebula and central star Ps 1 and K648 respectively may have
  been produced from a blue staggler. However, this interpretation is
  in doubt since the nebular mass of Ps1 is too small to be explained
  by this model. The high CSPN mass favors this hypothesis but the low
  nebular mass does not favor it. Additional mass-loss due to binary
  interaction could explain this.}
\item{This model suggests that there should be, in globular clusters
  with different populations with different helium abundances, two
  white dwarf cooling tracks with the second generation star with
  higher $Y$ producing a slightly more massive WD then the first
  generation star with a lower $Y$.}
\end{enumerate}

\section*{Acknowledgments}

\label{lastpage}

\end{document}